\documentclass{osa-article}
\journal{oe}

\usepackage{upgreek}

\usepackage{enumerate}
\usepackage{amsmath}
\newtheorem{thm}{Theorem}
\newtheorem{lem}{Lemma}

\usepackage{booktabs}

\begin{document}

\title{Single-scatter channel impulse response model of non-line-of-sight ultraviolet communications}

\author{Tian Cao,\authormark{1} Shihan Chen,\authormark{1} Tianfeng Wu,\authormark{1} Changyong Pan,\authormark{1,2} and Jian Song\authormark{1,2,*}}

\address{\authormark{1}Beijing National Research Center for Information Science and Technology (BNRist), Department of Electronic Engineering, Tsinghua University, Beijing 100084, China\\
\authormark{2}Key Laboratory of Digital TV System of Guangdong Province and Shenzhen City, Research Institute of Tsinghua University in Shenzhen, Shenzhen 518057, China}

\email{\authormark{*}jsong@tsinghua.edu.cn} 



\begin{abstract}
Previous studies on the temporal characteristics of single-scatter transmission in non-line-of-sight (NLOS) ultraviolet communications (UVC) were based on the prolate-spheroidal coordinate system. In this work, a novel single-scatter channel impulse response (CIR) model is proposed in the spherical coordinate system, which is more natural and comprehensible than the prolate-spheroidal coordinate system in practical applications. Additionally, the results of the widely accepted Monte-Carlo (MC)-based channel model of NLOS UVC are provided to verify the proposed single-scatter CIR model. Results indicate that the computational time costed by the proposed single-scatter CIR model is decreased to less than 0.7\% of the MC-based one with comparable accuracy in assessing the temporal characteristics of NLOS UVC channels.
\end{abstract}

\section{Introduction}
Ultraviolet communications (UVC) exploiting the strong scattering effect of solar-blind ultraviolet light caused by atmospheric molecules and aerosols can realize non-line-of-sight (NLOS) transmission\cite{2019_Vavoulas_ICST, 2022_Cao_OE, 2019_Song_OE}. NLOS UVC link can easily bypass obstacles and even achieve omnidirectional transmission and reception\cite{2021_Wu_OL, 2019_Li_OE}. That is, the alignment between transceivers is not essentially required in NLOS UVC systems, unlike other optical wireless communications (OWC), such as visible light communications and free-space optical communications\cite{2021_Cao_JQE,2018_Kaymak_ICST}. This merit can reduce the complexity of practical OWC systems. 

However, the ultraviolet light is scattered one or more times when it transmits from a light source to a detector via NLOS links, which makes building an analytical channel model of NLOS UVC knotty. In order to assess the characteristics of NLOS UVC links, numerical channel models of NLOS UVC based on Monte-Carlo (MC) methods with multiple scattering events considered have been proposed\cite{2011_Drost_JOSA,2020_Yuan_TC,2009_Ding_JSAC}. MC-based NLOS UVC channel models can obtain both path loss (PL) and channel impulse response (CIR), which are critical indicators of wireless channels. In addition, MC-based NLOS UVC channel models were experimentally validated and have been widely accepted by current studies\cite{2011_Drost_JOSA, 2009_Ding_JSAC, 2010_Ding_conf, 2012_Zuo_OE, 2011_Elshimy_JOSA, 2019_Wu_CL, 2021_Cao_OL}. Nevertheless, those models are time-consuming because they must generate and simulate a considerable number of photons to attain stable results. Fortunately, NLOS UVC is mainly designed for short-range communications where the distance between transceivers is up to tens of meters\cite{2019_Vavoulas_ICST}. The total received optical energy of ultraviolet light in NLOS UVC is dominated by that from the first scattering event under short-range conditions. Based on that fact, single-scatter channel models of NLOS UVC have attracted widespread attention in recent years because they can achieve almost the same results as MC-based ones with much less computational complexity\cite{2021_Cao_JQE, 2012_Zuo_OE, 2011_Elshimy_JOSA, 2019_Wu_CL, 2021_Cao_OL, 1991_Luettgen_JOSA, 2015_Wang_COL, 2016_Sun_JOSA}. 

The single-scatter channel models of NLOS UVC were developed in the prolate-spheroidal coordinate system in the early stage\cite{2011_Elshimy_JOSA, 1991_Luettgen_JOSA}. In those models, a transmitter (Tx) and a receiver (Rx) are placed at two focal points of a prolate-spheroidal coordinate system, respectively. Using the fact that the sum of the distance from each focal point to a prolate-spheroid surface is constant, the CIR of NLOS UVC links can be obtained. By accumulating the CIR results over time, the PL of  NLOS UVC can be further got. However, not only in MC-based channel models but also in practical experiments, the geometry of the NLOS UVC systems is described by the inclination (or elevation) and azimuth angles in the spherical coordinate system\cite{2011_Drost_JOSA, 2012_Zuo_OE}. Therefore, it is more natural and intelligible to investigate the single-scatter channel models of NLOS UVC in the spherical coordinate system.
Although some single-scatter PL models of  NLOS UVC in the spherical coordinate system have been increasingly reported very recently\cite{2021_Cao_JQE, 2012_Zuo_OE, 2019_Wu_CL, 2021_Cao_OL, 2015_Wang_COL, 2013_Zuo_OL}, only \cite{2019_Song_OE} and \cite{2017_Song_OC} approximately investigated the temporal characteristics of the  single-scatter NLOS UVC links based on the Riemann sum method in the spherical coordinate system. The exact single-scatter CIR model of  NLOS UVC in the spherical coordinate system has not been built so far.

Motivated by the above investigation, a novel single-scatter CIR model of NLOS UVC is proposed in the spherical coordinate system. Specifically, we establish the relationship between the time that a photon needs to travel from the Tx to the Rx via a single-scatter event and the radial distance of the spherical coordinate system at first. Then the expression of the total received optical energy at a given time is obtained, which is essentially a cumulative distribution function (CDF) of received optical energy in the time dimension. By differentiating that CDF with respect to time, the expression of the single-scatter CIR of NLOS UVC can be derived. Finally, the proposed single-scatter CIR model is verified by the widely accepted MC-based channel model of NLOS UVC.

\begin{figure}[!t]
	\centering
	\includegraphics[width=8cm]{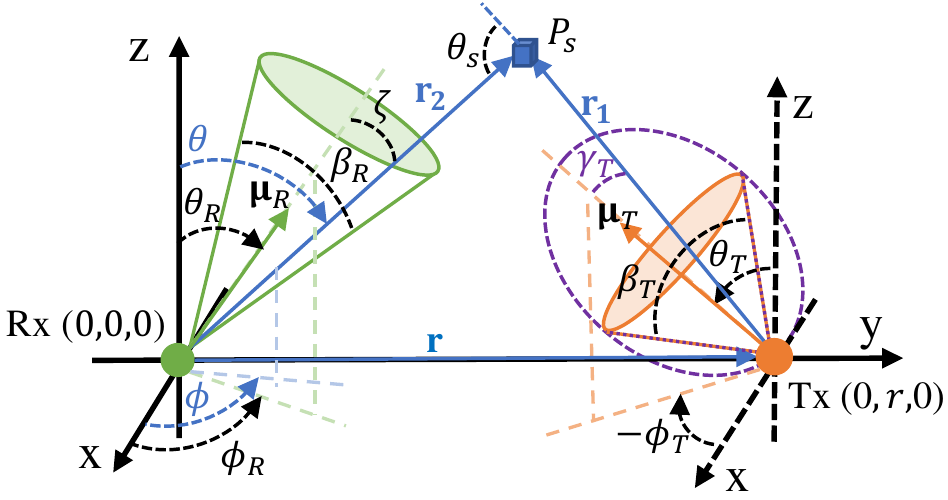}
	\caption{Diagram of a single-scatter NLOS UVC link.}
	\label{fig1}
\end{figure}

\section{System model and single-scatter channel impulse response analysis}
\label{sec2}

A single-scatter NLOS UVC link is illustrated in Fig. \ref{fig1}. A Rx and a Tx are placed at the origin and the point of $(0,r,0)$, respectively. The vector from the Rx to the Tx can be written as ${\bf{r}} = {\left[ {0,r,0} \right]^{\rm{T}}}$, where the superscript ``T'' denotes the transpose operator. ${P_s}$ denotes an arbitrary single-scatter event. The vectors from the Tx and the Rx to the ${P_s}$ are represented by ${{\bf{r}}_1}$ and ${{\bf{r}}_2}$, respectively. Here, coordinates $\theta $, $\phi $, and ${r_2} = \left\| {{{\bf{r}}_2}} \right\|$ are the inclination angle, azimuth angle, and the radial distance, respectively, of the spherical coordinate system utilized for the single-scatter CIR model, where $\left\|  \cdot  \right\|$ is the Euclidean norm. Therefore, we have ${{\bf{r}}_2} = {r_2}{\left[ {\sin \theta \cos \phi ,\sin \theta \sin \phi ,\cos \theta } \right]^{\rm{T}}}$ and ${{\bf{r}}_1} = {{\bf{r}}_2} - {\bf{r}}$. The Euclidean norm of ${{\bf{r}}_1}$ can be further obtained as
\begin{equation}
	{r_1} = \left\| {{{\bf{r}}_1}} \right\| = \sqrt {r_2^2 + {r^2} - 2{r_2}r\sin \theta \sin \phi } .
	\label{eq1}
\end{equation}

The field-of-view (FOV) of the Rx is conical with the apex angle equaling ${\beta _R}$, which also refers to the full FOV angle. The direction vector of the Rx’s FOV is represented by ${{\bf{\mu }}_R} = {\left[ {\sin {\theta _R}\cos {\phi _R},\sin {\theta _R}\sin {\phi _R},\cos {\theta _R}} \right]^{\rm{T}}}$, where $\theta _R$  and  ${\phi _R}$ are the inclination and the azimuth angles of the FOV’s axis, respectively. Let ${\cal T}\left( {{\gamma _T}} \right)$ be the emission pattern of the light source at the Tx, where ${\gamma _T} = \arccos \left[ {\frac{{\sin {\theta _T}\left( {{r_2}\sin \theta \cos \left( {\phi  - {\phi _T}} \right) - r\sin {\phi _T}} \right) + {r_2}\cos \theta \cos {\theta _T}}}{{{r_1}}}} \right]$ is the angle from the ${{\bf{\mu }}_T}$ to the ${{\bf{r}}_1}$, ${{\bf{\mu }}_T} = {\left[ {\sin {\theta _T}\cos {\phi _T},\sin {\theta _T}\sin {\phi _T},\cos {\theta _T}} \right]^{\rm{T}}}$ is the unit vector of the Tx’s pointing direction, ${\theta _T}$ and ${\phi _T}$ are the inclination and the azimuth angles of the Tx beam’s axis, respectively. ${\cal T}\left( {{\gamma _T}} \right)$ specifies the distribution of the emitted optical power per unit solid angle and refers to the emission pattern of the light source.
${\cal T}\left( {{\gamma _T}} \right)$ is supposed to be selected based on the practically utilized light source. Two distributions are used in this work to characterize the emission pattern:
\begin{itemize}
\item[1)]  Uniform distribution (UD) \cite{2012_Zuo_OE, 1991_Luettgen_JOSA}. It is demonstrated by the orange solid line at the Tx in Fig. \ref{fig1}. UD is the most representative distribution to describe the emission pattern of the light source in the field of NLOS UVC at present. It can be expressed as
\begin{equation}
	{\cal T}\left( {{\gamma _T}} \right) = \left\{ {\begin{array}{*{20}{c}}
			{\displaystyle\frac{1}{{2\pi \left( {1 - \cos \displaystyle\frac{{{\beta _T}}}{2}} \right)}},}&{0 \le {\gamma _T} \le \displaystyle \frac{{{\beta _T}}}{2}}\\
			{0,}&{\displaystyle \frac{{{\beta _T}}}{2} < {\gamma _T} \le \pi }
	\end{array}} \right.
	\label{eq2}
\end{equation}
where  ${\beta _T}$ denotes the beam divergence angle for the UD. 

\item[2)] Lambertian distribution (LD) \cite{2021_Cao_OL, 2012_Ghassemlooy_book}. It is illustrated by the purple dashed line at the Tx in Fig. \ref{fig1}. LD is recognized as the emission pattern of light-emitting diodes (LEDs), which are widely adopted in short-range NLOS UVC links in recent years. It can be written as
\begin{equation}
	{\cal T}\left( {{\gamma _T}} \right) = \left\{ {\begin{array}{*{20}{c}}
			{\displaystyle\frac{{m + 1}}{{2\pi }}{{\cos }^m}{\gamma _T},}&{0 \le {\gamma _T} \le \displaystyle\frac{\pi }{2}}\\
			{0,}&{\displaystyle\frac{\pi }{2} < {\gamma _T} \le \pi }
	\end{array}} \right.
	\label{eq3}
\end{equation}
where ${\beta _T}$ is the full angle at the half of the maximum intensity for the LD, $m = \frac{{ - \ln 2}}{{\ln \cos \frac{{{\beta _T}}}{2}}}$ is the order of Lambertian emission.

\end{itemize}

When an optical impulse is transmitted at time 0 with the energy of  $ Q_T $, the total received optical energy via a single-scatter link of NLOS UVC can be expressed as \cite{2021_Cao_OL}
\begin{equation}
	{Q_R} = {Q_T}{k_s}{A_r}\int_{{\theta _{\min }}}^{{\theta _{\max }}} {\int_{{\phi _{\min }}}^{{\phi _{\max }}} {\int_{{r_{{2_{\min }}}}}^{{r_{{2_{\max }}}}} {\frac{{{\cal T}\left( {{\gamma _T}} \right)\exp \left[ { - {k_e}\left( {{r_1} + {r_2}} \right)} \right]\cos \zeta {\mathop{\rm P}\nolimits} \left( {\cos {\theta _s}} \right)}}{{r_1^2}}\sin \theta {\mathop{\rm d}\nolimits} \theta {\mathop{\rm d}\nolimits} \phi {\mathop{\rm d}\nolimits} {r_2}} } } ,
	\label{eq4}
\end{equation}
where ${k_e} = {k_a} + {k_s}$ is the extinction coefficient, ${k_s}$ is the scattering coefficient, $ {k_a} $ is the absorption coefficient, ${A_r}$ is the detection area of the detector at the Rx, $\zeta  = \arccos \left[ \sin {\theta _R}\sin \theta \cos \left( \phi  -  \right. \right.$ $\left. \left. \phi _R  \right)  + \cos {\theta _R}\cos \theta  \right]$ is the angle between ${{\bf{\mu }}_R}$ and ${{\bf{r}}_2}$, ${\theta _s} = \arccos \left( {\frac{{r\sin \theta \sin \phi  - {r_2}}}{{{r_1}}}} \right)$ is the scattering angle between ${{\bf{r}}_1}$ and $ - {{\bf{r}}_2}$, and ${\mathop{\rm P}\nolimits} \left( {\cos {\theta _s}} \right)$ is the scattering phase function, which has been explained and given in Eq. (4) of \cite{2021_Cao_OL}. In addition, the calculations of upper and lower limits of the triple integral in \eqref{eq4}, that is, $\left[ {{\theta _{\min }},{\theta _{\max }}} \right]$, $\left[ {{\phi _{\min }},{\phi _{\max }}} \right]$, and $\left[ {{r_{{2_{\min }}}},{r_{{2_{\max }}}}} \right]$, can refer to \cite{2012_Zuo_OE} and \cite{2021_Cao_OL} for the emission pattern of light source obeying UD and LD, respectively.

We noticed that the CIR of the single-scatter NLOS UVC channel is intrinsically a probability density function (PDF), which indicates how much optical energy the Rx can receive during an infinitesimal time interval when an optical impulse is transmitted at time 0. Additionally, the total received optical energy from time 0 to $t$, which is essentially a CDF of received optical energy, depends on the length of the single-scatter trajectory, i.e., ${r_1} + {r_2}$, given $\theta $ and $\phi $. Therefore, we will derive the total received optical energy from time 0 to $t$ at first. By differentiating it with respect to time, the CIR of the single-scatter NLOS UVC channel in the spherical coordinate system can be further achieved. The above process will be given in detail below.

Let ${r_a}$ be the length of the single-scatter trajectory, i.e., ${r_a} = {r_1} + {r_2}$, and 
\begin{equation}
\tilde t\left( {{r_2}} \right) = \frac{{{r_a}}}{c} = \frac{{\sqrt {r_2^2 + {r^2} - 2{r_2}r\sin \theta \sin \phi }  + {r_2}}}{c},
	\label{eq5}
\end{equation}
be the time costed by the single-scatter trajectory, where $c$  is the light speed. 

\begin{lem} 
	\label{lem1}
	$\tilde t\left( {{r_2}} \right)$ is an increasing function of ${r_2}$ given $\theta $ and $\phi $.
\end{lem}
{\bf Proof \quad} 
In terms of Eq. \eqref{eq1}, $r_a$  can be expressed as
\begin{equation}
	{r_a} = \sqrt {r_2^2 + {r^2} - 2{r_2}r\sin \theta \sin \phi }  + {r_2}
	\label{eq6}
\end{equation}
By differentiating Eq. \eqref{eq6} with respect to  $r_2$, we have
\begin{equation}
	\frac{{{\mathop{\rm d}\nolimits} {r_a}}}{{{\mathop{\rm d}\nolimits} {r_2}}} = \frac{{{r_2} - r\sin \theta \sin \phi }}{{\sqrt {r_2^2 + {r^2} - 2{r_2}r\sin \theta \sin \phi } }} + 1.
	\label{eq7}
\end{equation}
We noticed that
\begin{equation}
\begin{array}{l}
	{\left[ {\displaystyle\frac{{{r_2} - r\sin \theta \sin \phi }}{{\sqrt {r_2^2 + {r^2} - 2{r_2}r\sin \theta \sin \phi } }}} \right]^2} = \displaystyle\frac{{r_2^2 + {r^2}{{\sin }^2}\theta {{\sin }^2}\phi  - 2{r_2}r\sin \theta \sin \phi }}{{r_2^2 + {r^2} - 2{r_2}r\sin \theta \sin \phi }} \le 1\\
	\Rightarrow  - 1 \le \displaystyle\frac{{{r_2} - r\sin \theta \sin \phi }}{{\sqrt {r_2^2 + {r^2} - 2{r_2}r\sin \theta \sin \phi } }} \le 1
\end{array}
	\label{eq8}
\end{equation}
because ${r^2}{\sin ^2}\theta {\sin ^2}\phi  \le {r^2}$. Applying the fact given in Eq. \eqref{eq8} to \eqref{eq7}, it can be obtained that $\frac{{{\mathop{\rm d}\nolimits} {r_a}}}{{{\mathop{\rm d}\nolimits} {r_2}}} \ge 0$. That is, ${r_a}$ increases as ${r_2}$ increases given $\theta $ and $\phi $. Thus, in accordance with Eq. \eqref{eq5}, Lemma \ref{lem1} can be achieved.  $\hfill\blacksquare$ 

Based on the Lemma \ref{lem1}, the shortest and longest time that the single-scatter trajectory costs is determined by the ${r_{{2_{\min }}}}$ and ${r_{{2_{\max }}}}$, respectively, given $\theta $ and $\phi $. Therefore, the total received optical energy from time 0 to $t$, i.e., the CDF, can be obtained by adjusting the interval of integration for ${r_2}$ as follows:
\begin{equation}
		\begin{aligned}
{F_{{Q_R}}}\left( t \right) &= {Q_T}{k_s}{A_r}\\
& \times \int_{{\theta _{\min }}}^{{\theta _{\max }}} {\int_{{\phi _{\min }}}^{{\phi _{\max }}} {\int_{{{\hat r}_{{2_{\min }}}}\left( t \right)}^{{{\hat r}_{{2_{\max }}}}\left( t \right)} {\frac{{{\cal T}\left( {{\gamma _T}} \right)\exp \left[ { - {k_e}\left( {{r_1} + {r_2}} \right)} \right]\cos \zeta {\mathop{\rm P}\nolimits} \left( {\cos {\theta _s}} \right)}}{{r_1^2}}\sin \theta {\mathop{\rm d}\nolimits} \theta {\mathop{\rm d}\nolimits} \phi {\mathop{\rm d}\nolimits} {r_2}} } } 
	\end{aligned},
	\label{eq9}
\end{equation}
where the upper bound and the lower bound of integration for $r_2$ are
\begin{equation}
{\hat r_{{2_{\max }}}}\left( t \right) = \left\{ {\begin{array}{*{20}{c}}
		{{r_{{2_{\max }}}},}&{t > \tilde t\left( {{r_{{2_{\max }}}}} \right)}\\
		{{{\tilde r}_2}(t),}&{t \le \tilde t\left( {{r_{{2_{\max }}}}} \right)}
\end{array}} \right.
	\label{eq10}
\end{equation}
and
\begin{equation}
{\hat r_{{2_{\min }}}}\left( t \right) = \left\{ {\begin{array}{*{20}{c}}
		{{r_{{2_{\min }}}},}&{t > \tilde t\left( {{r_{{2_{\min }}}}} \right)}\\
		{{{\tilde r}_2}(t),}&{t \le \tilde t\left( {{r_{{2_{\min }}}}} \right)}
\end{array}} \right. ,
	\label{eq11}
\end{equation}
respectively. ${\tilde r_2}(t) = \frac{{{{\left( {ct} \right)}^2} - {r^2}}}{{2ct - 2r\sin \theta \sin \phi }}$ is the value of ${r_2}$ when $\tilde t\left( {{r_2}} \right) = t$ and can be obtained by solving the following equation: 
\begin{equation}
\sqrt {\tilde r_2^2(t) + {r^2} - 2{{\tilde r}_2}(t)r\sin \theta \sin \phi }  + {\tilde r_2}(t) = ct.
	\label{eq12}
\end{equation}
It should be noted that ${\tilde r_2}(t)$ of interest locates within the interval of integration for ${r_2}$. This is because if $t \le \tilde t\left( {{r_{{2_{\min }}}}} \right)$, then the interval of integration for ${r_2}$ is $\left[ {{{\tilde r}_2}(t),{{\tilde r}_2}(t)} \right]$ and the integral equals zero; if $t > \tilde t\left( {{r_{{2_{\max }}}}} \right)$, then the interval of integration for ${r_2}$ is $\left[ {{r_{{2_{\min }}}},{r_{{2_{\max }}}}} \right]$, where ${\tilde r_2}(t)$ is not involved. 

The derivative  of ${F_{{Q_R}}}\left( t \right)$ with respect to time and subsequent normalization by the detector area yield the CIR of the single-scatter NLOS UVC channel with the unit of W/m$^2$, as given in the following Theorem.

\begin{thm}
	\label{thm1}
	The CIR of the single-scatter NLOS UVC channel in the spherical coordinate system can be expressed as
	\begin{equation}
		h\left( t \right) = {Q_T}{k_s}\int_{{\theta _{\min }}}^{{\theta _{\max }}} {\int_{{\phi _{\min }}}^{{\phi _{\max }}} {\varphi \left( {t,\theta ,\phi } \right){\mathop{\rm d}\nolimits} \theta {\mathop{\rm d}\nolimits} \phi } }  ,
		\label{eq13}
	\end{equation}
where 
	\begin{equation}
\varphi \left( {t,\theta ,\phi } \right) = \left\{ {\begin{array}{*{20}{c}}
		{\displaystyle\frac{{{\cal T}\left( {{\gamma _T}} \right)\exp \left[ { - {k_e}\left( {{r_1} + {{\tilde r}_2}(t)} \right)} \right]\cos \zeta {\mathop{\rm P}\nolimits} \left( {\cos {\theta _s}} \right)\sin \theta }}{{r_1^2}}{{\tilde r'}_2}(t),}&{\tilde t\left( {{r_{{2_{\min }}}}} \right) < t \le \tilde t\left( {{r_{{2_{\max }}}}} \right){\rm{ }}}\\
		{0,}&{{\rm{otherwise}}}
\end{array}} \right.
	\label{eq14}
\end{equation}
and ${\tilde r'_2}(t) = \frac{c}{{\left( {1 + \frac{{{{\tilde r}_2}(t) - r\sin \theta \sin \phi }}{{\sqrt {\tilde r_2^2(t) + {r^2} - 2{{\tilde r}_2}(t)r\sin \theta \sin \phi } }}} \right)}}$ is the first-order derivative of ${\tilde r_2}(t)$.
\end{thm}
{\bf Proof \quad} 
Using the relationship between PDF and CDF, we have
\begin{equation}
		\begin{aligned}
	h\left( t \right) &= \displaystyle\frac{{\displaystyle\frac{{{\mathop{\rm d}\nolimits} {F_{{Q_R}}}\left( t \right)}}{{{\mathop{\rm d}\nolimits} t}}}}{{{A_r}}}\\
	&= {Q_T}{k_s}\left\{ {\int_{{\theta _{\min }}}^{{\theta _{\max }}} {\int_{{\phi _{\min }}}^{{\phi _{\max }}} {\displaystyle\frac{{{\cal T}\left( {{\gamma _T}} \right)\exp \left[ { - {k_e}\left( {{r_1} + {{\hat r}_{{2_{\max }}}}\left( t \right)} \right)} \right]\cos \zeta {\mathop{\rm P}\nolimits} \left( {\cos {\theta _s}} \right)\sin \theta }}{{r_1^2}}\displaystyle\frac{{{\mathop{\rm d}\nolimits} {{\hat r}_{{2_{\max }}}}\left( t \right)}}{{{\mathop{\rm d}\nolimits} t}}{\mathop{\rm d}\nolimits} \theta {\mathop{\rm d}\nolimits} \phi } } } \right.\\
	&- \int_{{\theta _{\min }}}^{{\theta _{\max }}} {\int_{{\phi _{\min }}}^{{\phi _{\max }}} {\displaystyle\frac{{{\cal T}\left( {{\gamma _T}} \right)\exp \left[ { - {k_e}\left( {{r_1} + {{\hat r}_{{2_{\min }}}}\left( t \right)} \right)} \right]\cos \zeta {\mathop{\rm P}\nolimits} \left( {\cos {\theta _s}} \right)\sin \theta }}{{r_1^2}}} } \left. {\displaystyle\frac{{{\mathop{\rm d}\nolimits} {{\hat r}_{{2_{\min }}}}\left( t \right)}}{{{\mathop{\rm d}\nolimits} t}}{\mathop{\rm d}\nolimits} \theta {\mathop{\rm d}\nolimits} \phi } \right\}
\end{aligned}
	\label{eq15}
\end{equation}

According to Eq (10), $\frac{{{\mathop{\rm d}\nolimits} {{\hat r}_{{2_{\max }}}}\left( t \right)}}{{{\mathop{\rm d}\nolimits} t}} = {\tilde r'_2}(t)$ if $t \le \tilde t\left( {{r_{{2_{\max }}}}} \right)$, otherwise $\frac{{{\mathop{\rm d}\nolimits} {{\hat r}_{{2_{\max }}}}\left( t \right)}}{{{\mathop{\rm d}\nolimits} t}} = 0$. Similarly, based on Eq. \eqref{eq11}, $\frac{{{\mathop{\rm d}\nolimits} {{\hat r}_{{2_{\min }}}}\left( t \right)}}{{{\mathop{\rm d}\nolimits} t}} = {\tilde r'_2}(t)$ if $t \le \tilde t\left( {{r_{{2_{\min }}}}} \right)$, otherwise $\frac{{{\mathop{\rm d}\nolimits} {{\hat r}_{{2_{\min }}}}\left( t \right)}}{{{\mathop{\rm d}\nolimits} t}} = 0$. In addition, applying the technique of implicit differentiation to Eq. \eqref{eq12}, the first-order derivative of ${\tilde r_2}(t)$ can be obtained as 
\begin{equation}
\begin{array}{l}
	\displaystyle\frac{{{\mathop{\rm d}\nolimits} \left[ {\sqrt {\tilde r_2^2(t) + {r^2} - 2{{\tilde r}_2}(t)r\sin \theta \sin \phi }  + {{\tilde r}_2}(t)} \right]}}{{{\mathop{\rm d}\nolimits} t}} = \displaystyle\frac{{{\mathop{\rm d}\nolimits} \left( {ct} \right)}}{{{\mathop{\rm d}\nolimits} t}}\\
	\Rightarrow \displaystyle\frac{{{\mathop{\rm d}\nolimits} {{\tilde r}_2}(t)}}{{{\mathop{\rm d}\nolimits} t}} = \displaystyle\frac{c}{{\left( {1 + \frac{{{{\tilde r}_2}(t) - r\sin \theta \sin \phi }}{{\sqrt {\tilde r_2^2(t) + {r^2} - 2{{\tilde r}_2}(t)r\sin \theta \sin \phi } }}} \right)}}
\end{array}
	\label{eq16}
\end{equation}
Substituting  $\frac{{{\mathop{\rm d}\nolimits} {{\hat r}_{{2_{\max }}}}\left( t \right)}}{{{\mathop{\rm d}\nolimits} t}}$,   $\frac{{{\mathop{\rm d}\nolimits} {{\hat r}_{{2_{\min }}}}\left( t \right)}}{{{\mathop{\rm d}\nolimits} t}}$, and Eq. \eqref{eq16} into Eq. \eqref{eq15}, we can achieve Theorem \ref{thm1}. $\hfill\blacksquare$

\section{Results and discussion}
\label{sec3}

\begin{figure}[!bh]
	\centering
	\includegraphics[width=8cm]{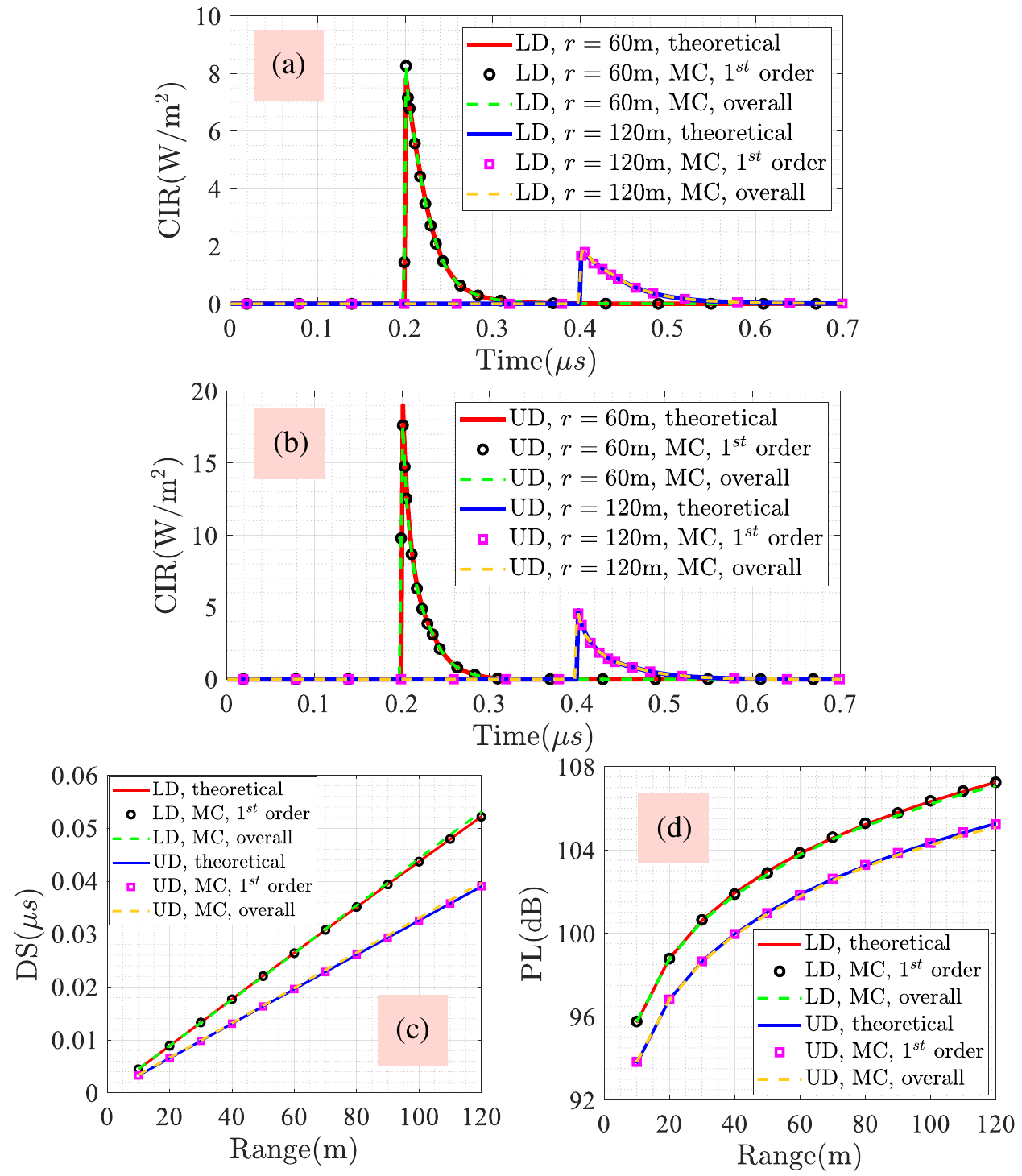}
	\caption{The CIR of NLOS UVC links when the emission pattern of the light source obeys (a) LD and (b) UD with both $r = 60$ m and $r = 120$ m considered. (c) DS and (d) PL of the NLOS UVC links versus the baseline range between the Tx and the Rx.}
	\label{fig2}
\end{figure}

The CIR results of NLOS UVC computed through the proposed single-scatter CIR model are demonstrated and discussed in this section. The emission patterns of the light sources obeying UD and LD are both considered. Additionally, the delay spread (DS) is also provided to quantitatively evaluate the temporal characteristics of NLOS UVC channels and defined as \cite{2017_Sun_PJ, 2015_EL_JLT}
\begin{equation}
DS = {\left[ {\frac{{\int {{{\left( {t - \mu } \right)}^2}h\left( t \right){\mathop{\rm d}\nolimits} t} }}{{\int {h\left( t \right){\mathop{\rm d}\nolimits} t} }}} \right]^{\frac{1}{2}}},
	\label{eq17}
\end{equation}
where $\mu  = {{\int {th\left( t \right){\mathop{\rm d}\nolimits} t} } \mathord{\left/
		{\vphantom {{\int {th\left( t \right){\mathop{\rm d}\nolimits} t} } {\int {h\left( t \right){\mathop{\rm d}\nolimits} t} }}} \right.
		\kern-\nulldelimiterspace} {\int {h\left( t \right){\mathop{\rm d}\nolimits} t} }}$. A large DS means a low coherence bandwidth of the wireless channel. The results of PL, another important indicator of NLOS UVC channels, are given here and can be expressed in decibel (dB) as follows:
\begin{equation}
PL =  - 10{\log _{10}}\frac{{{Q_R}}}{{{Q_T}}}.
	\label{eq18}
\end{equation}
Besides, the results of CIR and PL obtained through the MC-based NLOS UVC channel model \cite{2011_Drost_JOSA} with the first-order scattering events (labeled ``1st order'' in legends) and multiple scattering events (labeled ``overall'' in legends) considered are exhibited to verify the corresponding theoretical results. The scattering order in the MC-based channel model is set to 3 under the multiple scattering condition. Geometric parameters of NLOS UVC systems and atmospheric parameters used in the computation are extracted from \cite{2021_Cao_JQE, 2021_Cao_OL} and listed as follows unless otherwise specified: ${A_r} = 1.92 \ {\rm{c}}{{\rm{m}}^2}$, ${\beta _R} = {30^ \circ }$, ${\beta _T} = {60^ \circ }$, ${\phi _T} =  - {90^ \circ }$, ${\phi _R} = {90^ \circ }$, ${Q_T} = 1$ J, $\gamma  = 0.017$, $g = 0.72$, $f = 0.5$, Rayleigh scattering coefficient ${k_{s,r}} = 0.24 \ {\rm{k}}{{\rm{m}}^{ - 1}}$, Mie scattering coefficient ${k_{s,m}} = 0.25 \ {\rm{k}}{{\rm{m}}^{ - 1}}$, and ${k_a} = 0.9 \ {\rm{k}}{{\rm{m}}^{ - 1}}$, where $\gamma $, $g$, and $f$ are model parameters in ${\mathop{\rm P}\nolimits} \left( {\cos {\theta _s}} \right)$, ${k_s} = {k_{s,r}} + {k_{s,m}}$.

The influence of changing the range between the Tx and the Rx on the channel characteristics of NLOS UVC is shown in Fig. \ref{fig2}. Here, we set ${\theta _T} = {\theta _R} = {60^ \circ }$. The CIR of NLOS UVC links with the emission pattern of the light source obeying UD and LD are given in Fig. \ref{fig2} (a) and (b), respectively. Theoretical CIR results obtained through the proposed single-scatter CIR model demonstrate good agreement with those got by the MC-based channel model with both first-order and multiple scattering events considered because the total received optical energy is dominated by that from the first-order scattering event, which verifies the proposed single-scatter CIR model. Additionally, for both emission patterns, the CIR curve of $r = 60$ m rises earlier than that of $r = 120$ m. This is because the longer the baseline range is, the more time it takes for the light to travel from the Tx to the Rx. The DS and PL of NLOS UVC links versus the baseline range are further given in Fig. \ref{fig2} (c) and (d), respectively. It can be found from Fig. \ref{fig2} (c) that with the increase of baseline range between the Tx and the Rx, the DS of NLOS UVC links increases. The coherence bandwidth of the NLOS UVC channel thus becomes narrower. This is because as the baseline range increases, the common volume between the beam and the FOV where single-scatter events occur becomes larger, which makes the possible time interval that a photon arrives at the Rx wider. It can be seen from Fig. \ref{fig2} (d) that with the increase of the baseline range, PL increases, i.e., the received optical energy decreases. Therefore, when the baseline range increases, the CIR curve becomes shorter and broader, and the area under it becomes less, as the CIR results shown in Fig. \ref{fig2} (a) and (b). That is, the channel condition of NLOS UVC goes worse with the increase of the baseline range between the Tx and the Rx. 

\begin{figure}[!h]
	\centering
	\includegraphics[width=8cm]{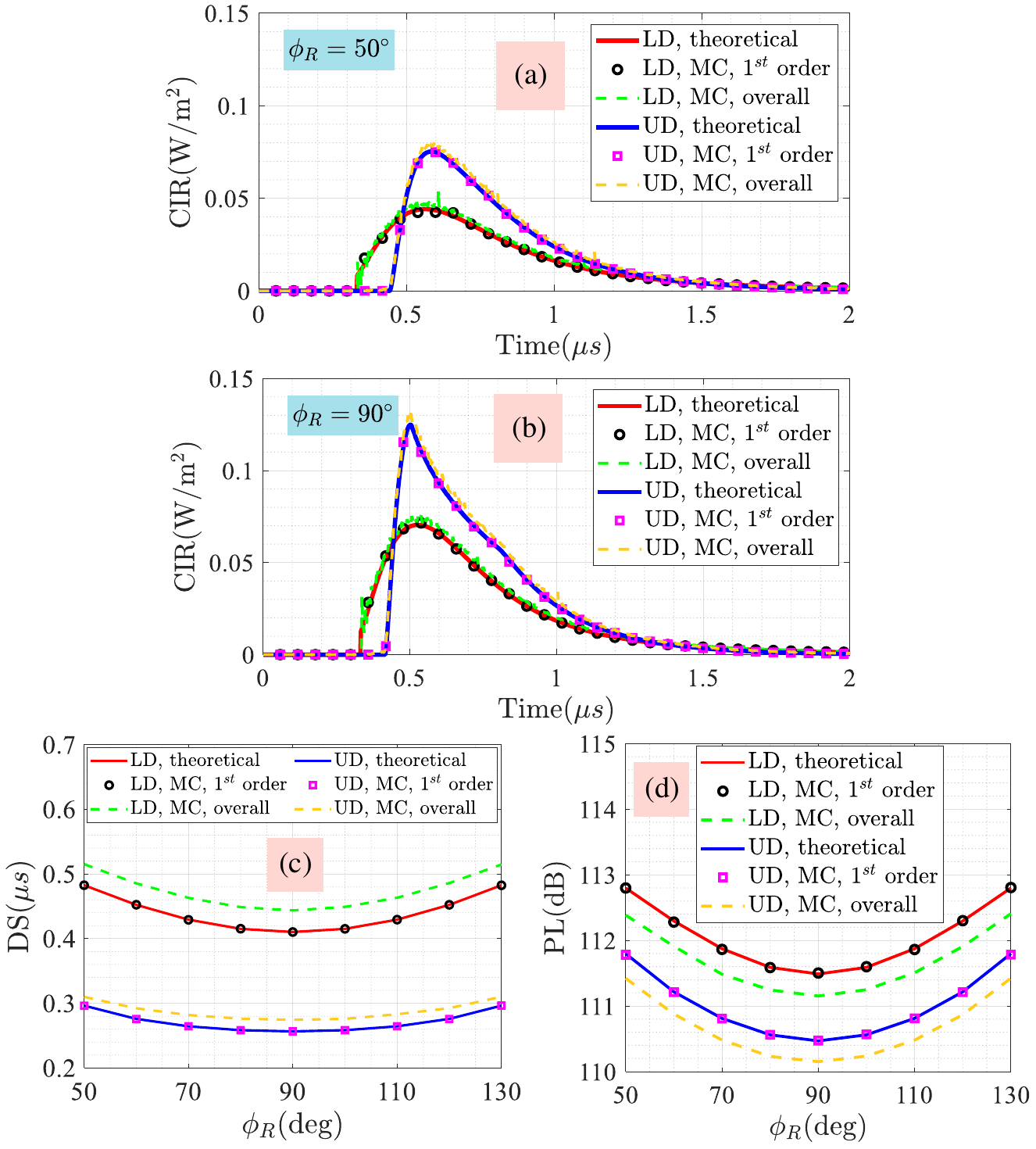}
	\caption{The CIR of NLOS UVC links with (a) ${\phi _R} = {50^ \circ }$ and (b) ${\phi _R} = {90^ \circ }$. (c) DS and (d) PL of the NLOS UVC links versus ${\phi _R}$.}
	\label{fig3}
\end{figure}
Fig. \ref{fig3} illustrates the relationship between channel characteristics of NLOS UVC and ${\phi _R}$. Specifically, the CIR results of NLOS UVC links with ${\phi _R}$ equaling ${50^ \circ }$ and ${90^ \circ }$ are given in Fig. \ref{fig3} (a) and (b), respectively. Fig. \ref{fig3} (c) and (d) plot the DS and PL results of NLOS UVC links, respectively, versus ${\phi _R}$. Here, we set $r = 100$ m and ${\theta _T} = {\theta _R} = {30^ \circ }$. It can be found that the theoretical single-scatter CIR results in Fig. \ref{fig3} (a) and (b) match well with the corresponding CIR results obtained by the MC-based channel model with the first-order scattering events only considered. However, the overall MC curves are jagged and slightly higher than single-scatter ones. This is because considering multiple scattering events increases the randomness of discrete photon arrivals in the MC-based channel model and broadens the time interval that a photon can arrive at the Rx. Hence the DS of the overall CIR results obtained by the MC-based channel model is also slightly larger than that of single-scatter CIR results, as shown in Fig. \ref{fig3} (c). At the same time, more optical energy can be received by the Rx via higher-order scattering events, leading to lower PL, as shown in Fig. \ref{fig3} (d). However, the difference between the CIR curves in Fig. \ref{fig3} (a) and (b) of the single-scatter and multiple-scatter cases is unobvious and acceptable in practice. It is worth noting that the computational complexity of the proposed single-scatter CIR model is much less than that of the MC-based one. Specifically, to obtain Fig. \ref{fig3} (a), the proposed single-scatter CIR model costs about 6.8 s whereas the MC-based one costs 1225.8 s on a laptop with a CPU of 2.8 GHz and memory of 16 GB, where we set the temporal resolution to 2 ns and the number of photons in the MC-based channel model to ${10^8}$. The computational time costed by the proposed single-scatter CIR model is reduced by about 3 orders of magnitude compared with the MC-based one. In addition, it can be seen from Fig. \ref{fig3} (c) and (d) that both DS and PL have a minimum value when ${\phi _R}$ equals ${90^ \circ }$. This is because when ${\phi _R}$ equals ${90^ \circ }$, the NLOS UVC link is in coplanar geometry, where the axes of the beam and FOV are in the same plane. Besides, comparing Figs. \ref{fig2} and \ref{fig3} shows that increasing the inclination angle could improve the channel condition of NLOS UVC links. For example, when ${\theta _T} = {\theta _R} = {30^ \circ }$ and ${\phi _R} = {90^ \circ }$, the single-scatter DS and PL for the emission pattern obeying LD are roughly 0.41 $\mu {\rm{s}}$ and 111.5 dB, respectively, as shown in Fig. \ref{fig3} (c) and (d). However, when ${\theta _T} = {\theta _R} = {60^ \circ }$ and $r = 100$ m, the single-scatter DS and PL for the emission pattern obeying LD are about 0.044 $\mu {\rm{s}}$ and 106 dB, respectively, as shown in Fig. \ref{fig2} (c) and (d), which manifests an obvious improvement in both DS and PL.

\begin{figure}[!h]
	\centering
	\includegraphics[width=8cm]{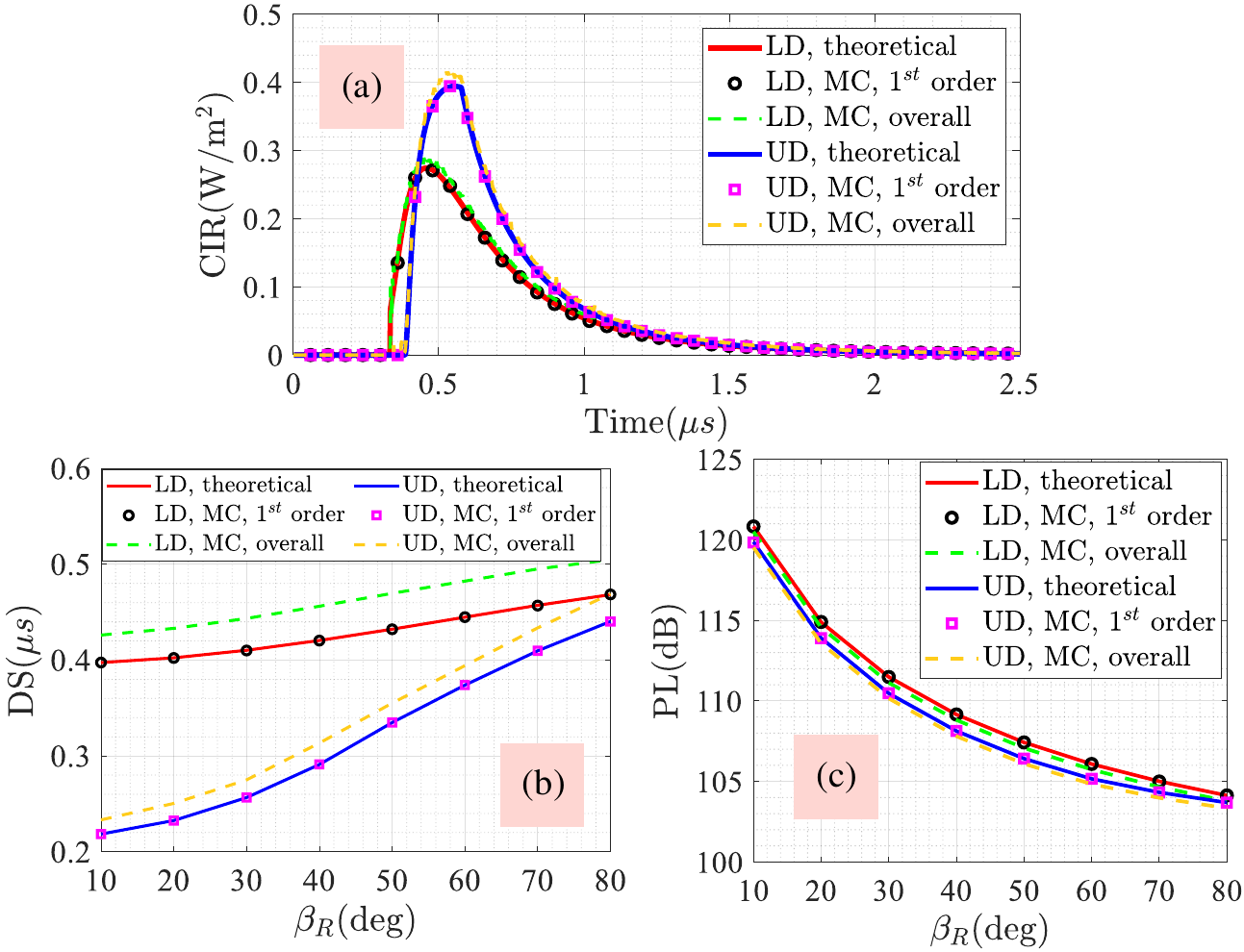}
	\caption{(a) The CIR of NLOS UVC links with ${\beta _R} = {60^ \circ }$. (b) DS and (c) PL of the NLOS UVC links versus ${\beta _R}$.}
	\label{fig4}
\end{figure}
Fig. \ref{fig4} presents the influences of changing ${\beta _R}$ on the channel characteristics of NLOS UVC. Fig. \ref{fig4} (a) gives the CIR curves of NLOS UVC links when ${\beta _R}$ equals ${60^ \circ }$. DS and PL of NLOS UVC links against ${\beta _R}$ are depicted in Fig. \ref{fig4} (b) and (c), respectively. Here, we set $r = 100$ m and ${\theta _T} = {\theta _R} = {30^ \circ }$. It can be found from Fig. \ref{fig4} (b) and (c) that DS increases, but PL decreases, as ${\beta _R}$ increases. That is, although a large full FOV angle is beneficial to increase the received optical energy and further reduce the PL, it would also enlarge the common volume between the beam and the FOV, increase DS, and hence reduce the coherence bandwidth. For example, comparing Figs. \ref{fig3} (b) and \ref{fig4} (a), it can be seen that the CIR curves in Fig. \ref{fig4} (a) are slightly broader and the area under them is larger than those in the Figs. \ref{fig3} (b). In addition, from Fig. \ref{fig4} (b) we can further observe that the influence of increasing ${\beta _R}$ on the DS performance for the light source with the uniformly distributed emission pattern is more severe than that for the light source with the emission pattern obeying LD. For example, from ${\beta _R} = 10$ to ${\beta _R} = 80$, the DS of the single-scatter NLOS UVC link with the emission pattern of light source obeying LD increases by 17.1\%, whereas that obeying UD increases by 101.8\%. This is because the light radiation is uniformly distributed within the beam divergence angle and suddenly disappears when ${\gamma _T}$ is greater than ${\beta _T}$ for the uniformly distributed emission pattern. Under this condition, changing ${\beta _R}$ could obviously vary the common volume between the beam and the FOV, leading to a distinct change in DS. However, for the light source with the emission pattern obeying LD, the light radiation is distributed within a hemisphere and the intensity of light smoothly reduces to zero from the center axis to the margin of the beam. Thus, changing ${\beta _R}$ exerts a greater impact on the DS performance with the emission pattern of the light source obeying UD than obeying LD in the considered NLOS UVC system.

\section{Conclusion}
In this work, a novel single-scatter CIR model of NLOS UVC was proposed based on the spherical coordinate system, which makes the investigation of the NLOS UVC channel more comprehensible and user-friendly than the prolate-spheroidal coordinate system. The validity of the proposed single-scatter CIR model of NLOS UVC was verified by the widely accepted MC-based NLOS UVC channel model. The results demonstrate that the proposed single-scatter CIR model of NLOS UVC could cost much less computational time than the MC-based one. In addition, reducing the baseline range between the Tx and Rx or adjusting the geometry of the NLOS UVC system to the coplanar case can reduce both the DS and the PL. Nevertheless, widening the full FOV angle can reduce the PL but increase the DS. This work will be helpful for analyzing the temporal characteristics of the NLOS UVC systems.

\begin{backmatter}
	\bmsection{Funding} Science, Technology and Innovation Commission of Shenzhen Municipality (NO. JSGG2021102 9095003004).
	\bmsection{Disclosures} The authors declare no conflicts of interest.
	\bmsection{Data availability} Data underlying the results presented in this paper are not publicly available at this time but may be obtained from the authors upon reasonable request.
\end{backmatter}


\bibliography{myReferences}

\end{document}